\documentclass[final,3p,times,12pt]{elsarticle}

\usepackage{amsmath,amsfonts}
\usepackage{algorithmic}
\usepackage{algorithm}
\usepackage{array}
\usepackage[caption=false,font=normalsize,labelfont=sf,textfont=sf]{subfig}
\usepackage{textcomp}
\usepackage{stfloats}
\usepackage{url}
\usepackage{xcolor}
\usepackage{graphicx,graphics}
\usepackage{epsfig,comment,hyperref,verbatim}
\begin{document}

\begin{frontmatter}
\journal{Computer Speech and Language}

\title{{Investigation of Self-supervised Pre-trained Models for Classification of Voice Quality from Speech and Neck Surface Accelerometer Signals}}

\author{Sudarsana Reddy Kadiri, Farhad Javanmardi, Paavo Alku}
\address{Department of Information and Communications Engineering, Aalto University, Finland}

\cortext[mycorrespondingauthor]{Corresponding author (Sudarsana Reddy Kadiri; sudarsana.kadiri@aalto.fi)}

\begin{abstract}
Prior studies in the automatic classification of voice quality have mainly studied the use of the acoustic speech signal as input. Recently, a few studies have been carried out by jointly using both speech and neck surface accelerometer (NSA) signals as inputs, and by extracting mel-frequency cepstral coefficients (MFCCs) and glottal source features. This study examines simultaneously-recorded speech and NSA signals in the classification of voice quality (breathy, modal, and pressed) using features derived from three self-supervised pre-trained models (wav2vec2-BASE, wav2vec2-LARGE, and HuBERT) and using a support vector machine (SVM) as well as convolutional neural networks (CNNs) as classifiers. Furthermore, the effectiveness of the pre-trained models is compared in feature extraction between glottal source waveforms and raw signal waveforms for both speech and NSA inputs. Using two signal processing methods (quasi-closed phase (QCP) glottal inverse filtering and zero frequency filtering (ZFF)), glottal source waveforms are estimated from both speech and NSA signals. The study has three main goals: (1) to study whether features derived from pre-trained models improve classification accuracy compared to conventional features {(spectrogram, mel-spectrogram, MFCCs, i-vector, and x-vector)}, (2) to investigate which of the two modalities (speech $vs.$ NSA) is more effective as input in the classification task with pre-trained model-based features, and (3) to evaluate whether the deep learning-based CNN classifier can enhance the classification accuracy in comparison to the SVM classifier. The results revealed that the use of the NSA input showed better classification performance compared to the speech signal. Between the features, the pre-trained model-based features showed better classification accuracies, both for speech and NSA inputs compared to the conventional features. The two classifiers performed equally well for all the pre-trained model-based features for both speech and NSA signals. It was also found that the HuBERT features performed better than the wav2vec2-BASE and wav2vec2-LARGE features for both speech and NSA inputs. In particular, when compared to the conventional features, the HuBERT features showed an absolute accuracy improvement of { 3\%--6\%}  for speech and NSA signals in the classification of voice quality. 
\end{abstract}

\end{frontmatter}

\section{Introduction}
Humans have the ability to produce different phonation types, such as breathy, tense/pressed, creaky, and falsetto, by controlling the activation of laryngeal muscles in conjunction with respiratory effort \cite{Lavar80,VQ_ASP,pietrowicz2017acoustic}. Phonation type is closely linked to voice quality, which is described as the auditory characteristics of a speaker's voice \cite{Lavar80,titze2000principles,Gordon02}. Breathy and tense voices are typically regarded as the two extremes of the voice quality continuum. Voice quality plays a significant role in conveying paralinguistic information, such as vocal emotions and personality in speech \cite{vq_4d,vq_perception1,vq_affect,vq_depression,birkholz2015contribution}. For example, breathy voice is commonly used to express politeness and intimacy \cite{politeness}, while pressed voice is used to convey emotions of high arousal, such as anger, disgust, and excitement \cite{vq_F0cues,SC43}. Moreover, voice quality is used in some languages to generate phonological contrasts  \cite{Gordon02,phonation_lang,kuang2014vocal,esposito2010effects,khan2012phonetics}. Further information about functioning of the speech production mechanism in humans and its role in generating voice qualities can be found in \cite{Lavar80,titze2000principles}.

Modal phonation is often used as a reference point in comparing different voice qualities. In modal phonation, laryngeal tension is typically low, and vocal fold vibration is mostly periodic. 
Moreover, vocal fold closure usually happens abruptly in modal voices. In contrast, the production of breathy voices involves weaker laryngeal tension, and closure of the glottis is  typically partial, resulting in the generation of turbulent noise. This results in spectra that show more prominent low-frequency harmonic components for breathy voices compared to modal voices. On the other hand, the production of pressed voices typically involves increased adduction and longitudinal tension of the vocal folds together with sharper glottal closure, resulting in stronger high-frequency harmonics compared to modal voices. This study aims to investigate the classification of voice quality into the above three classes of phonation: breathy, modal, and pressed.

Because voice quality is closely related to the vibration pattern of the vocal folds, changes in voice quality will affect the characteristics of the acoustical excitation waveform generated by the vocal folds, the glottal flow pulse \cite{Lavar80,titze2000principles,vq_compare}. Research has shown that the shape of the glottal pulse, estimated by glottal inverse filtering, changes from a smooth, almost symmetrical waveform in breathy voice to an asymmetrical pulse in pressed voice \cite{vq_compare,effect_F0}. These changes in the time-domain glottal pulse waveform are reflected in the frequency domain by spectral decay of the voice source \cite{LFSE,hillenbrand1994acoustic}. Different features have been proposed to analyze properties of the glottal source waveform, both in the time domain and frequency domain  \cite{vq_compare,MDQ,borsky2017modal}. Examples of time-domain features include the closing quotient, the speed quotient, and the quasi-open quotient \cite{vq_compare,Paavo11,Drugman-GSP}. Amplitude-based features (e.g., the normalized amplitude quotient) have also been proposed to parameterize the glottal source \cite{vq_compare,Paavo11,Drugman-GSP}. In the frequency domain, several measures (e.g., the amplitude difference between the fundamental and the second harmonic, and the harmonic richness factor) have been used to quantify the decay of the voice source spectrum \cite{VQ_ASP,hillenbrand1994acoustic,PSP}. 
The maximum dispersion quotient feature, which is based on the residual signal of linear prediction, was studied in \cite{MDQ} to capture sharper changes in vocal fold closure in pressed voices. Other features, such as the fundamental frequency, the harmonic ratio, and the spectral slope, were used to estimate the effect  of the glottal source on the speech spectrum in \cite{vq_affect,spectralslope,Kreiman12,vq_variability}.  In \cite{MDQ,borsky2017modal,phonation_speech,ph_speech_singing}, features of the glottal source (based on the linear prediction residual and the zero frequency filtered (ZFF) signal) were combined with mel-frequency cepstral coefficients (MFCCs) to classify voice qualities from speech and singing voices. Other features analyzed in \cite{phonation_speech,ph_speech_singing} include the energy of the ZFF signal, the slope of the ZFF signal at zero crossings, the energy of excitation, and the loudness measure. Moreover, cepstral coefficients derived from the spectra, computed by single frequency filtering and zero-time windowing, were studied in \cite{kadiri2019mel} in voice quality classification from speech and singing. {In \cite{brandner2023classification}, modulation power spectral features were studied in classification of voice quality in classical singing. Recently, voice quality detection (i.e., the detection of 
different voice qualities in each singing file, along with their onset and offset times for each detected voice quality) was studied in \cite{wang2023phonation}. 
The authors proposed an end-to-end encoder-decoder model as an alternative to the popular pipeline approach (i.e., a system consisting of a separate feature extraction stage followed by a classifier stage) for predicting voice quality labels along with their boundaries.}

The NSA sensor captures vibrations of the vocal folds during speech production, and it has been found to be useful in measuring vocal fold vibration characteristics \cite{stevens1975miniature,rendon2007mapping,mehta2015using}. As the NSA is mounted on the skin surface below the glottis, data collected by NSA sensors is less affected by the vocal tract resonances in comparison to acoustic speech data collected by microphones. This makes the NSA signal a valuable modality to analyze and classify voice qualities, as this signal provides insight into the vibration dynamics of the vocal folds. In a few previous studies, the NSA signal has been taken advantage of as a supplement to the acoustic speech signal in investigating vocal dose measures and hyperfunctional voice production \cite{mehta2016relationships,dose_nsa,coleman1988comparison,ghassemi2014learning}. However, only a few studies have investigated automatic classification of voice qualities based on NSA signals. In these studies, different techniques, such as the Gaussian mixture model, linear discriminant analysis, decision tree, support vector machine, K-nearest neighbours, and multi-layer perceptron classifiers, have been used in combination with features (e.g., harmonics, jitter, shimmer, entropy, and MFCCs) to classify voice qualities, such as breathy, modal, pressed, and rough \cite{borsky2017classification,lei2019discrimination}.

{To the best of our knowledge, automatic voice quality classification based on $simultaneous$ recordings of acoustic speech signals and NSA signals has only been studied in three investigations
\cite{kadiri2021glottal,kadiri2022convolutional,wlodarczak2022classification}. In \cite{kadiri2021glottal}, a variety of glottal source features, along with MFCCs, were used to discriminate between three different voice qualities (breathy, modal, and pressed) from speech and NSA signals (collected from 31 speakers) using SVM as classifier. It was found that the combination of glottal source features and MFCCs gave improved classification accuracy compared to individual features. Using the same data, it was shown in \cite{kadiri2022convolutional} that convolutional neural networks (CNNs) improved the performance compared to the SVM classifier when the classifiers used the spectrogram and mel-spectrogram features. In \cite{wlodarczak2022classification}, three voice qualities (breathy, neutral, pressed) were included but the authors investigated only two binary classification tasks (breathy vs. neutral and neutral vs. pressed) and reported that the discrimination between neutral and breathy voice lead to a lower accuracy compared to the discrimination between neutral and pressed voice.

It is worth noting that all the three previous studies referred to above are based on using conventional hand-crafted features. These conventional features require domain expertise, limiting their generalization and adaptability to different tasks. In contrast, $pre-trained$ $model$ $features$ learned from large-scale data offer better generalization capabilities, and save time and resources in feature engineering for various down-stream tasks.  Pre-trained model features are computed based on self-supervised representation learning that has become a topic of increasing interest in the field of para-linguistics, as most databases in this area are relatively small compared to the databases in automatic speech recognition (ASR). Self-supervised representation learning corresponds to training a model in an unsupervised manner using large databases. The model learns representations from the raw speech, and these representations can be used in the required downstream task. Examples of pre-trained models are wav2vec2 and HuBERT that have shown good performance in various speech technology tasks, such as ASR, emotion recognition, speaker and language identification, and voice disorder detection~\cite{NEURIPS2020wav2vec2,hernandez2022cross,mohamed2021arabic,vaessen2022fine,fan2020exploring,ribas2023automatic,saska2023Pretrained}. There are, however, no studies on using recent self-supervised pre-trained models, such as wav2vec2 \cite{NEURIPS2020wav2vec2} and HuBERT~\cite{hsu2021hubert}, for voice quality classification.

\subsection{Goals of the study}
The main goal of the current study is to investigate the classification of three voice qualities (breathy, modal, and pressed) by extracting features using self-supervised pre-trained models (wav2vec2 and HuBERT). Two modalities (speech vs. NSA) are compared as the input to the feature extraction. In addition, both the speech and NSA input are represented in two forms: either as original, raw waveforms or as glottal source signals that are estimated by inverse filtering the raw waveform. The wav2vec2 and HuBERT pre-trained models used in the present study are available at HuggingFace \cite{wolf2019huggingface}.

The major contributions of this study are:
\begin{itemize}
    \item Conducting a layer-wise analysis of self-supervised pre-trained models (wav2vec2 and HuBERT embeddings as features) for the voice quality classification from speech and NSA signals.
    
    \item Conducting a systematic comparison between five conventional features (spectrogram, mel-spectrogram, MFCCs, i-vector, and x-vector) and the pre-trained model embeddings as features, using SVM and CNN classifiers.

    \item Studying the effectiveness of two methods of deriving glottal source waveforms (quasi-closed phase (QCP) glottal inverse filtering and zero frequency filtering) from speech and NSA signals.

    \item Investigating the classification abilities of features (both conventional and pre-trained model embeddings) derived from glottal source waveforms of speech and NSA signals for the voice quality classification.

\end{itemize}}

\section{Voice Quality Classification System}
\begin{figure*}[htp]
\centering
\includegraphics[width=0.95\textwidth,height=4.8cm]{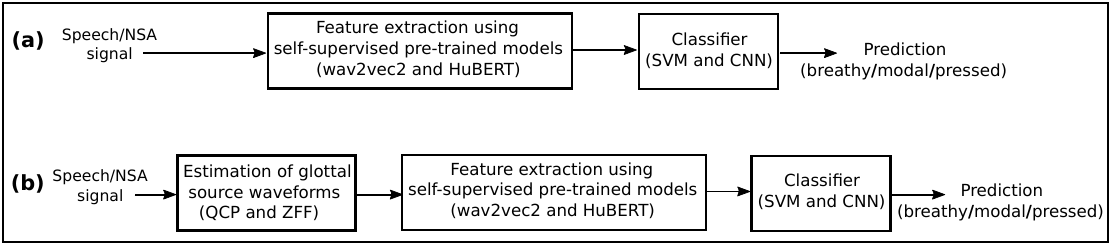}
\caption{\label{fig:VQframework} Schematic block diagrams describing voice quality classification systems based on the two-stage pipeline architecture. Two raw signals (speech and NSA) are used as input. In system (a), the raw signals are used directly as input to the feature extraction stage that is based on pre-trained models. In system (b), glottal source waveforms are estimated from the raw signals and used as input to the feature extraction stage. This study investigates glottal source waveforms derived using two signal processing methods (QCP and ZFF), two self-supervised pre-trained models (wav2vec2 and HuBERT), and two classifiers (SVM and CNN).}
\end{figure*}

Voice quality classification systems are built in this study using the popular two-stage pipeline approach consisting of a feature extraction stage and a classifier stage  as shown in Fig. \ref{fig:VQframework}. The feature extraction block uses three popular self-supervised pre-trained models (wav2vec2-BASE \cite{NEURIPS2020wav2vec2}, wav2vec2-LARGE\cite{NEURIPS2020wav2vec2}, and HuBERT \cite{hsu2021hubert}) to extract feature vectors from raw signals (speech and NSA) and from glottal source waveforms (estimated from speech and NSA). Two signal processing algorithms (quasi-closed phase (QCP) glottal inverse filtering  \cite{Airaksinen2014} and ZFF  \cite{Murty32}) are used to compute glottal source waveforms. In the classifier block, one machine learning (ML) -based classifier (support vector machine (SVM)) and one deep learning (DL) -based classifier (CNN) are used to predict the voice quality class. The descriptions of the signal processing methods, self-supervised pre-trained models and classifiers are given in the next sub-sections.

\subsection{Extraction of Glottal Source Waveforms}
\label{ssec:GSW}
In this section, two algorithms (QCP and ZFF), which are used for the estimation of glottal source in the current investigation, are described.
It is important to note that when the acoustic speech signal recorded by a microphone outside the mouth is used as input to inverse filtering, the aim is to cancel the effects of supraglottal resonances from the input. However, when inverse filtering is used for the NSA signal, the aim is to cancel the effects of subglottal resonances. The QCP and ZFF methods were originally developed for acoustic speech inputs, but they are used in the current investigation both for the acoustic speech input as well as for the NSA input to extract information about voice production. For the sake of clarity, we describe in the following these signal processing algorithms using a generic time-domain input, $s[n]$, which refers both to the acoustic speech signal and the NSA signal.

\subsubsection{Extraction of Glottal Source Waveforms using the QCP Method}
\begin{figure}[htbp]
\centering
\vspace{-0.1cm}
\includegraphics[width=0.9\columnwidth,height=3cm,trim={0.1cm 0cm 0cm 0cm},clip]{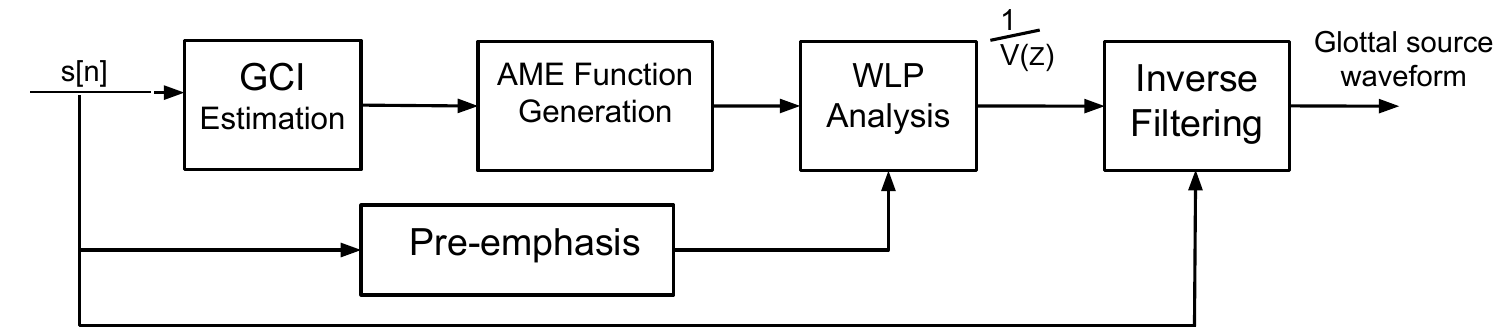}
\vspace{-0.2cm}
\caption{\label{qcpbd} Block diagram of the quasi-closed phase (QCP) glottal inverse filtering method.}
\end{figure}

The QCP method \cite{Airaksinen2014} (shown in Fig.~\ref{qcpbd}), is a glottal inverse filtering (GIF) method to estimate the glottal excitation. QCP is based on the principles of an older GIF method called closed phase analysis \cite{Wong79}. In closed phase analysis, the vocal tract model is estimated using linear prediction (LP) from a few individual speech samples located in a short interval after glottal closure. Unlike in closed phase analysis, however, all the samples of the analysis frame are taken advantage of in QCP to compute an autoregressive filter model for the vocal tract. This computation is done using a temporally weighted linear prediction (WLP) analysis that uses the attenuated main excitation (AME) \cite{AME} weighting function. The AME function is a temporal waveform, which de-emphasizes the square of the LP error signal at those time-instants where the effect of the glottal source is strong. As a result, the resulting all-pole WLP model (denoted by $V(z)$ in Fig.~\ref{qcpbd}) models the characteristics of the vocal tract better. The AME waveform is generated automatically based on three parameters (the duration quotient, the position quotient, and the level of the AME function at glottal closure), whose values can be pre-set to those proposed in \cite{Airaksinen2014}. To generate the AME waveform, glottal closure instants (GCIs) need to be extracted first. After the vocal tract model has been computed with WLP analysis, using the AME function, the input signal ($s[n]$) is inverse filtered with $V(z)$ to estimate the glottal source waveform. 

\subsubsection{Extraction of Glottal Source Waveforms using the ZFF Method}
The ZFF method was proposed in \cite{Murty32} as a way to analyze the glottal source characteristics, specifically at glottal closure instants. ZFF is based on the fact that the spectral effect of an impulse (which occurs, for example, at glottal closure) is present over the entire spectrum, including zero frequency (0 Hz). In the ZFF algorithm, the pre-emphasized signal ($x[n] = s[n] - s[n-1]$) is first filtered by a cascade of two zero frequency resonators (ZFRs). ZFR is a filter that has a double pole at $z$=1.0 in the $z$-plane. The filtering can be represented by the following equation:

\begin{equation}
z_o[n]=\sum_{k=1}^{4}{a_k z_o[n-k]} + x[n],
\end{equation}

where $a_1=+4$, $a_2=-6$, $a_3=+4$, $a_4=-1$. This process is equivalent to integrating or cumulatively summing the signal four times, which causes the signal to grow as a polynomial function of time. To remove this growing trend, the local mean is computed over the average pitch period and subtracted from $z_o[n]$. The resulting signal is called the zero frequency filtered signal and is given by:

\begin{equation}
z[n]=z_o[n]-\frac{1}{2M+1}\sum_{i=-M}^{M}{z_o[n+i]}.
\end{equation}

\begin{figure}[h]
\centering
 \vspace{-0.2cm}
\includegraphics[width=0.9\columnwidth,height=3cm,trim={0.1cm 0cm 0cm 0cm},clip]{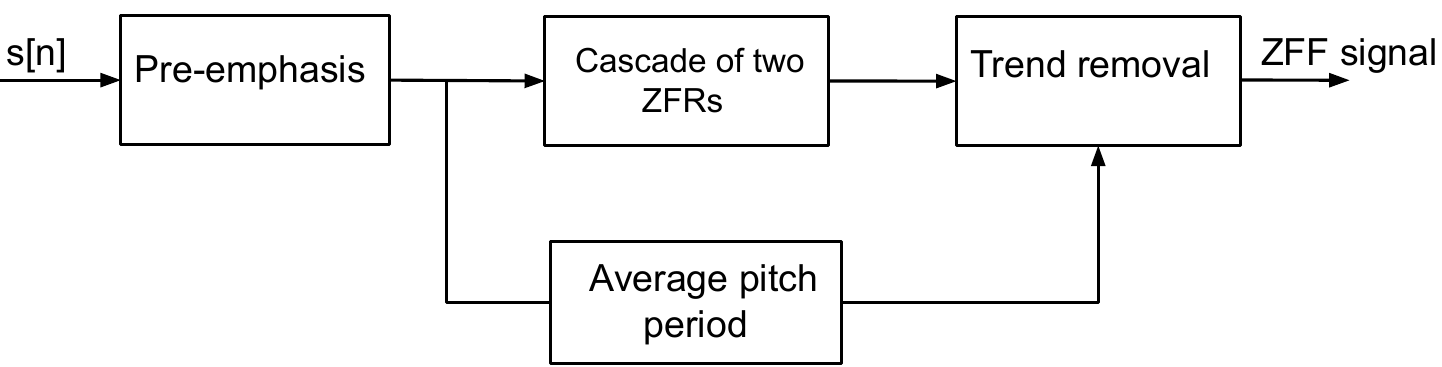}
\vspace{-.2cm}
 \caption{\label{zffbd}Block diagram of the zero frequency filtering (ZFF) method.}
\end{figure}

Here, $2M+1$ is the number of samples used to remove the trend. The ZFF signal is considered an approximate glottal source waveform \cite{kadiri2020analysis,kadiri2019mfcc-vsw}, and the negative-to-positive zero-crossings (NPZCs) correspond to GCIs when considering the positive polarity of the signal \cite{Murty32,myspd}. The steps involved in the ZFF method are shown in Fig.~\ref{zffbd}.

\begin{figure}[h]
\begin{center}
\includegraphics[width=0.9\columnwidth, height=8cm,trim={0.8cm 0cm 0.8cm 0.2cm},clip]{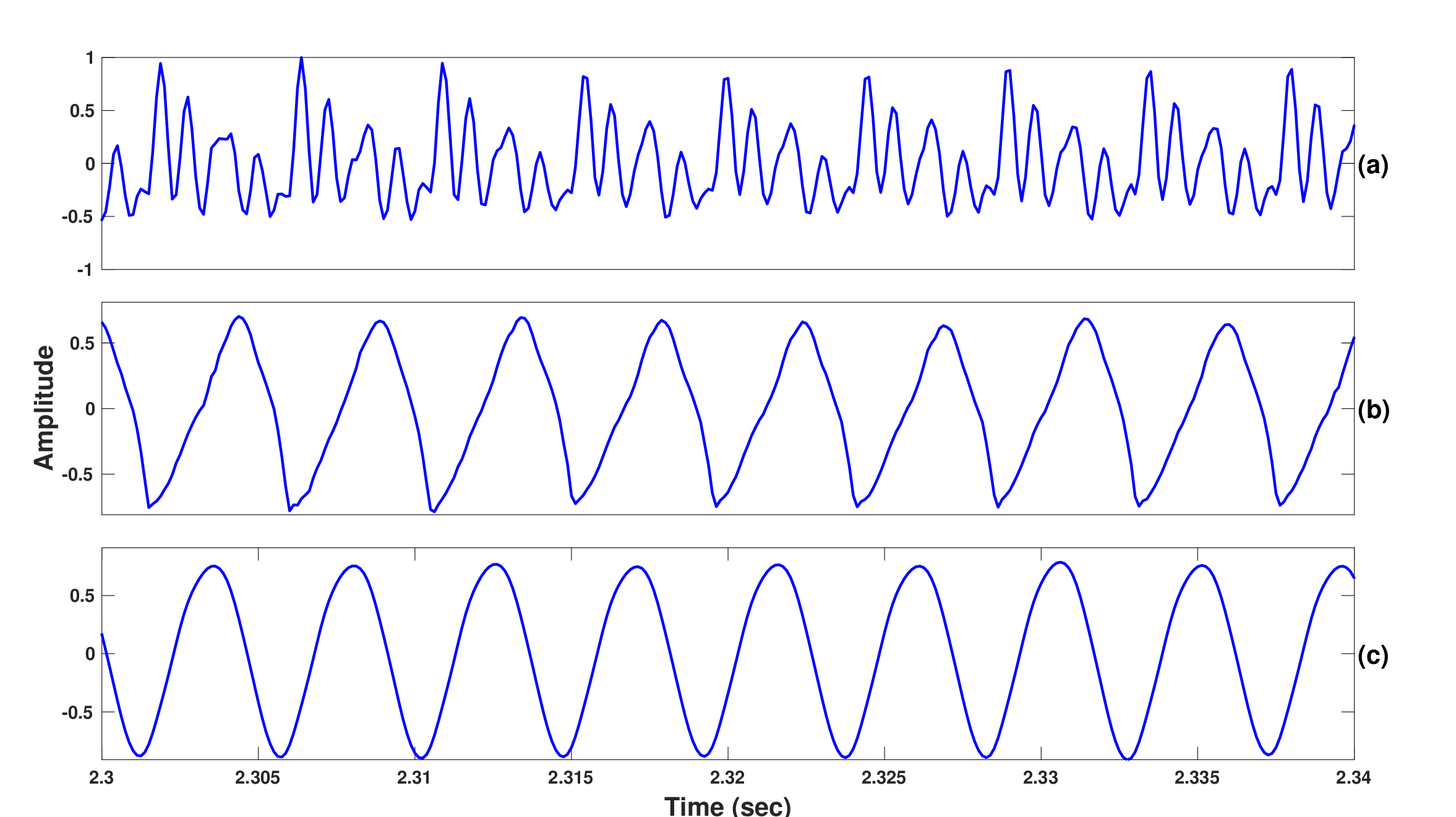}
\caption{Illustration of glottal source waveforms derived for speech signal (shown in (a)) using the QCP (shown in (b)) and ZFF (shown in (c)) methods.}

\label{mic_vsw}
\end{center}
\vspace{-0.5cm}
\end{figure}

\begin{figure}[h]
\begin{center}
\includegraphics[width=0.9\columnwidth, height=8cm,trim={0.8cm 0cm 0.8cm 0.2cm},clip]{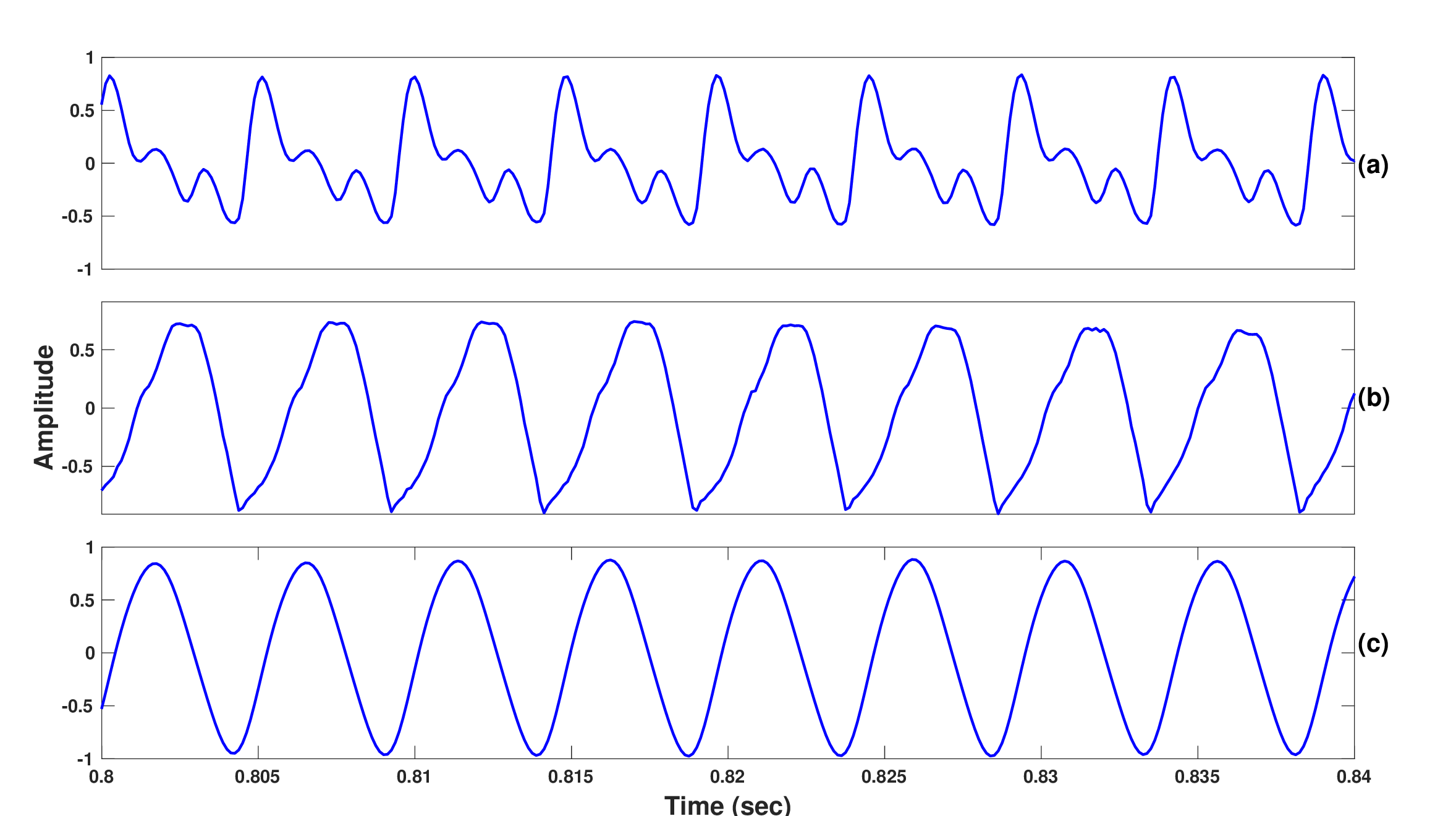}
\caption{Illustration of glottal source waveforms derived for NSA signal (shown in (a)) using the QCP (shown in (b)) and ZFF (shown in (c)) methods.}
\label{nsa_vsw}
\end{center}
\vspace{-0.5cm}

\end{figure}

\begin{figure}[htb]
\centering
\includegraphics[width=\columnwidth,height=7cm,trim={2.1cm 0cm 3.5cm 0cm},clip]{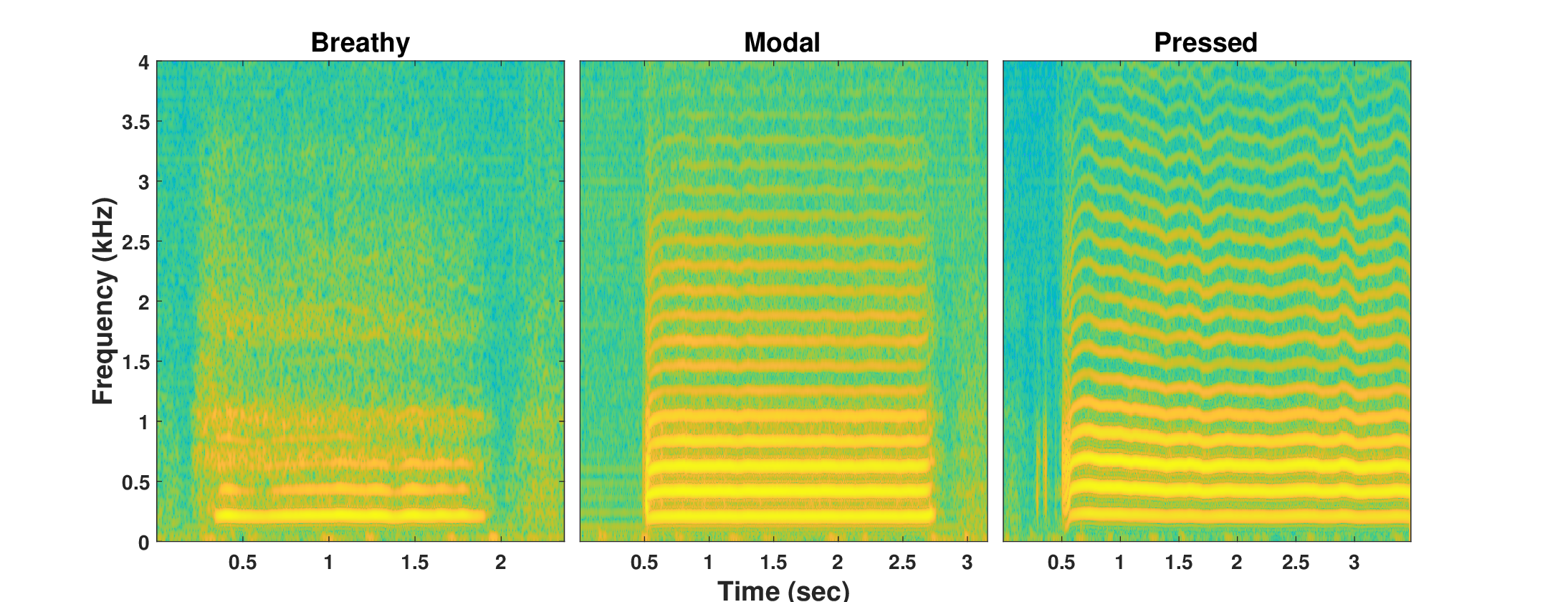}
\vspace{-0.5cm}
\caption{\label{spectrograms}Illustration of spectrograms for the speech signal of vowel [a] in breathy, modal and pressed voices.}
\end{figure}

To demonstrate glottal source waveforms estimated by the QCP and ZFF algorithms, a segment of acoustic speech signal, shown in Fig.~\ref{mic_vsw}, and the corresponding simultaneously recorded NSA signal, shown Fig.~\ref{nsa_vsw}, are considered. In both figures, panel (a) shows the input (i.e., speech vs. NSA), and panels (b) and (c) show the glottal source waveforms computed by QCP and ZFF, respectively. These figures demonstrate that the time-domain waveforms of the glottal source shown in panels (b) and (c) are much smoother and more elementary signals compared to the waveforms of the speech signal (shown in panel (a) of  Fig.~\ref{mic_vsw}) and the NSA signal (shown in panel (a) of Fig.~\ref{nsa_vsw}). Particularly, the waveform of the speech signal (panel (a) in Fig.~\ref{mic_vsw}) shows rapid fluctuations caused by resonances of the (supra-glottal) vocal tract. It can be seen in panels (b) and (c) of Fig.~\ref{mic_vsw} that these fluctuations have been removed by QCP and ZFF. For a second illustration, we show in Fig.~\ref{spectrograms} typical speech spectrograms for the three voice qualities that are investigated in this study. 
The spectrograms were computed from breathy, modal, and pressed utterances of the vowel [a]. It can be observed that the strength of the harmonics between low and high frequencies varies considerably between the voice qualities. In the remainder of the paper, the glottal source waveforms derived by the QCP method for speech and NSA signals are referred to as speech-QCP and NSA-QCP, respectively. Similarly, the approximate glottal source waveforms derived by the ZFF method for speech and NSA signals are referred to as speech-ZFF and NSA-ZFF, respectively.

\subsection{Feature Extraction Based on Pre-trained Models}
In this study, the effectiveness of features extracted using three pre-trained models in the classification of voice qualities is evaluated. The three pre-trained models utilized in feature extraction are wav2vec2-BASE \cite{NEURIPS2020wav2vec2}, wav2vec2-LARGE \cite{NEURIPS2020wav2vec2}, and HuBERT ~\cite{hsu2021hubert}. These models have been pre-trained on a large unsupervised audio dataset, and fine-tuned using a small dataset for ASR, allowing the final layers of the models to learn speech representations that contain phoneme-related information \cite{NEURIPS2020wav2vec2, fan2020exploring}. There is also information related to phones in the learned speech representations from the lower layers of the network. Therefore, the pre-trained models can be used to extract features in various speech-related tasks, such as in classification of stuttering and Alzheimer's disease, and in speaker and language identification ~\cite{mohamed2021arabic,vaessen2022fine,fan2020exploring,sheikh2022introducing, gauder2021alzheimer}.

\subsubsection{Wav2Vec2.0}
{Two wav2vec2 models, namely wav2vec2-BASE and wav2vec2-LARGE, are explored in the current study. The wav2vec2 model contains a CNN encoder, a context network, and a quantization module. The context network consists of 12 transformer blocks with a model dimension of 768 in wav2vec2-BASE, and of 24 transformer blocks with a model dimension of 1024 in wav2vec2-LARGE. In this study, the raw signals (speech and NSA) and their glottal source waveforms are fed as input to the wav2vec2 models. The temporal average of the inputs to the first transformer layer, together with the outputs of all transformer layers of the context network, are computed to extract a total of thirteen 768-dimensional feature vectors for the wav2vec2-BASE model, and a total of twenty-five 1024-dimensional feature vectors for the wav2vec2-LARGE model.}

\subsubsection{HuBERT}
{The HuBERT-LARGE model architecture consists of a CNN encoder, a context network (which contains 24 transformer layers), a projection layer, and a code embedding layer. In the current study, the temporal average of the inputs of the first transformer layer and the output of each of the 24 transformer layers of the context network are used as features in classification. For each signal (speech/NSA and their glottal source waveforms), a total of twenty-five 1024-dimensional feature vectors are extracted.} 

{The wav2vec2-BASE model consists of approximately 95M parameters, wav2vec2-LARGE consists of approximately 315M parameters, and HuBERT-LARGE consists of approximately  317M parameters.} In the remaining part of this article, we refer to the features extracted from the three pre-trained models described above as the wav2vec2-BASE, wav2vec2-LARGE, and HuBERT features.

\subsection{Classifiers}

In the current study, one ML classifier (SVM) and one DL classifier (CNN) are utilized to classify voice quality of the speech/NSA signal to three classes (breathy, modal, and pressed). Training of the SVM classifiers was performed with the radial basis function (RBF) kernel and a regularization parameter value of 1. In addition, the gamma parameter of SVM was fixed to have a value defined as \( \gamma = 1 / (D \cdot Var(X)) \), where $Var(X)$ is the variance of the training data, and D is the dimension of the feature vectors. The architecture of the CNN classifier consists of three sequential convolutional layers, a flattened layer, and two fully-connected (dense) layers. Each convolutional layer is followed by batch normalization and the ReLU activation function. The second dense layer with the softmax activation function performs the multi-class classification task. Table \ref{tab:CNN-arch} shows the architecture of the CNN classifier in more details. Training of CNN was carried out using the following hyper-parameters: 100 epochs {with 20 epochs} as the early stopping, Adam optimizer with a learning rate of 0.001, batch size of 64, and the cross entropy loss function. In this study, the SVM and CNN classifiers were implemented using the Scikit-learn library \cite{scikit-learn} and Pytorch library \cite{NEURIPS2019_9015}, respectively.

\begin{table}[h!]
\begin{center}
\caption{\label{tab:CNN-arch} CNN architecture used in this study. Here $Conv$ refers to convolution layer and $FC$ refers to fully connected layer.}
\vspace{-0.1cm}
\centering
\resizebox{\columnwidth}{!}{
\begin{tabular}{|l|c|c|c|c|c|c|} \hline
Layers &Conv1& Conv2& Conv3& Flatten& FC1& FC2 \\ \hline \hline
No. filters/output dim.& 8& 16& 32& 32*(\#feature dimension/8)&16 &3 \\\hline
Kernel size & 3&3 &3 &- & -&- \\\hline
Stride & 2& 2& 2&-&- &-\\
\hline 
\end{tabular}}
\vspace{-0.2cm}
\end{center}
\end{table}

\section{Experimental setup}
\subsection{Database}
In this study, we use a voice quality repository described in \cite{lei2019discrimination}. This database contains five Canadian English vowel sounds ($[a]$, $[æ]$, $[e]$, $[i]$, and $[u]$) that were pronounced using three voice qualities (breathy, modal, and pressed) \cite{lei2019discrimination}. Each utterance was repeated three times by 31 native female speakers between the ages of 18 and 40. In total, the database includes 1395 vowel utterances (5 vowels $\times$ 3 voice qualities $\times$ 3 repetitions $\times$ 31 speakers). The database includes recordings of both the speech and NSA signals. The original sampling frequency in the recordings was 44.1 kHz, but we down-sampled the data to 16 kHz for the purposes of the current study. The speech signals were evaluated by five speech language pathologists, and the scores were analyzed for both intra- and inter-rater reliability. Out of the 952 signals that were considered reliable in the experts' screening, there were 395 breathy voices, 285 modal voices, and 272 pressed voices. {The total duration of the speech/NSA data taken from the repository to the current study is approximately 52 minutes.} To the best of the authors' knowledge, the repository used is the only publicly open database on voice quality that includes simultaneously recorded speech and NSA signals.

\subsection{Baseline features used for comparison}
In order to compare the performance of the pre-trained model-based features (wav2vec2-BASE, wav2vec2-LARGE, and HuBERT) with conventional reference features, five popular features (spectrogram, mel-spectrogram, MFCCs, i-vector, and x-vector) were used. In  \cite{kadiri2021glottal,kadiri2022convolutional}, three spectral features (spectrogram, mel-spectrogram, and MFCCs) have shown promising results in the automatic classification of voice qualities. The computation of these conventional features is summarized below.

\subsubsection{Spectrogram}
The speech/NSA signal is Hamming-windowed into overlapping time-frames of 25 ms with a shift of 5 ms, and the amplitude spectrum is estimated using a 1024-point fast Fourier transform (FFT). Then, the spectrogram features are computed by taking the logarithm of the amplitude spectrum for all the time-frames. Finally, the spectrogram features are averaged over the time axis to yield a 513-dimensional feature vector per speech/NSA signal.

\subsubsection{Mel-Spectrogram}
In order to compute the mel-spectrogram features, a mel-filterbank of 80 filters is applied to the amplitude spectrum, and the resulting mel-spectrogram is converted to a decibel scale by using the logarithm. For each speech/NSA signal, a 80-dimensional feature vector is obtained by averaging the mel-spectrogram features over time axis.

\subsubsection{MFCCs}

The MFCCs features are computed by applying the discrete cosine transform (DCT) on the mel-scale spectrum. From the resulting mel-cepstrum, the first 13 coefficients are considered and their first and second derivatives are computed. By averaging the MFCCs features over the time axis, a 39-dimensional feature vector is yielded per speech/NSA signal.

{
\subsubsection{i-vector}
The motivation behind the extraction of i-vectors stems from factor analysis modeling, which represents features in terms of uncorrelated components \cite{dehak,kenny2008study}. In this approach, a Gaussian Mixture Model-Universal Background Model (GMM-UBM) is adapted to capture a variable-length utterance as a fixed representation known as the super-vector. Subsequently, through factor analysis, the super-vectors are compressed to retain only the uncorrelated low-dimensional components, resulting in i-vectors. Initially, i-vectors were introduced for speaker verification \cite{dehak}. Subsequently, they were experimented in  language identification \cite{dehak2011language} and dialect classification \cite{kethireddy2022exploration}. We employed the Kaldi's i-vector extractor (available in \url{http://www.kaldi-asr.org/models/m7}) and the dimension of the i-vector was 400 for each signal.

\subsubsection{x-vector}
The deep neural embeddings, also known as the x-vector, refer to the embeddings that are derived from a deep neural network (DNN) specifically trained for dialect classification. Initially, x-vectors were introduced for extracting speaker embeddings \cite{7846260}. Subsequently, they were used in language identification \cite{snyder2018spoken} and dialect classification \cite{hanani_naser_2020,jain2018improved,kethireddy2022exploration}. We employed the Kaldi's x-vector extractor (available in \url{http://www.kaldi-asr.org/models/m7}) and the dimension of the x-vector was 512 for each signal.

The conventional features described in the above paragraphs are derived using the raw speech/NSA signal as input. In addition to using these raw signals as input, the same features were computed using as input the glottal source signals that were estimated from the speech and NSA signals as described in Section~\ref{ssec:GSW}.} 

\subsection{Training, testing and evaluation}
The classification experiments were conducted using the leave-one-speaker-out (LOSO) cross-validation scheme, where the data of one speaker was considered as testing data and the data of the remaining 30 speakers were used to train both SVM and CNN classifiers. In each iteration, the training and testing data were normalized by using the mean and standard deviation of the training data, and the classification accuracy for each of the testing speaker was computed. By considering one speaker for testing, a total of 31 iterations (training and testing process) were performed. The final result (classification accuracy) was obtained by  averaging over all iterations. In addition to accuracy metrics, confusion matrices were computed for assessing the performance. It should be noted that for the classification experiments with the CNN classifier, the data of one speaker from training data was used as validation in each iteration. {Note that the total duration of the speech/NSA data is approximately 52 minutes. Hence, the training data size is approximately 50 minutes and the testing data size is approximately 2 minutes for each iteration.}

\section{Results and Discussion}
This section first reports the results of the classification experiments obtained using the conventional features derived from both speech and NSA signals for the SVM and CNN classifiers. After that, the results obtained using the features derived from the three pre-trained models (wav2vec2-BASE, wav2vec2-LARGE, and HuBERT) are discussed.

\begin{table*}[h]
\centering
\caption{\label{Tab:Result_nsa_speech_conventional} {Results for the five conventional features (in terms of the mean and standard deviation of accuracy) for speech and NSA signals using the SVM and CNN classifiers.}}
\vspace{0.1cm}
\begin{tabular}{|l|ll||ll|}
\hline
\multicolumn{1}{|c|}{{\color[HTML]{000000} }}                           & \multicolumn{2}{c||}{{\color[HTML]{000000} SVM}}                                                              & \multicolumn{2}{c|}{{\color[HTML]{000000} CNN}}                                                            \\ \cline{2-5} 
\multicolumn{1}{|c|}{{{\color[HTML]{000000} Feature}}} & \multicolumn{1}{c|}{{\color[HTML]{000000} Speech}}         & \multicolumn{1}{c||}{{\color[HTML]{000000} NSA}} & \multicolumn{1}{c|}{{\color[HTML]{000000} Speech}}       & \multicolumn{1}{c|}{{\color[HTML]{000000} NSA}} \\ \hline \hline
{\color[HTML]{000000} Spectrogram}                                      & \multicolumn{1}{l|}{{\color[HTML]{000000} 81.72$\pm$6.09}} & {\color[HTML]{000000} 91.38$\pm$4.74}             & \multicolumn{1}{l|}{{\color[HTML]{000000} 86.18$\pm$6.58}} & {\color[HTML]{000000} 92.19$\pm$5.86}             \\ \hline
{\color[HTML]{000000} Mel-spectrogram}                                  & \multicolumn{1}{l|}{{\color[HTML]{000000} {83.99}$\pm$6.73}}   & {\color[HTML]{000000} {\bf92.84}$\pm$4.99}             & \multicolumn{1}{l|}{{\color[HTML]{000000} {\bf87.74}$\pm$5.65}} & {\color[HTML]{000000} {\bf92.76}$\pm$4.84}             \\ \hline
{\color[HTML]{000000} MFCC}                                             & \multicolumn{1}{l|}{{\color[HTML]{000000} 79.78$\pm$6.79}}   & {\color[HTML]{000000} 89.01$\pm$5.09}             & \multicolumn{1}{l|}{{\color[HTML]{000000} 77.93$\pm$6.41}} & {\color[HTML]{000000} 87.36$\pm$4.65}             \\ \hline
{\color[HTML]{000000}  i-vector}                                  & \multicolumn{1}{l|}{{\color[HTML]{000000} 81.05$\pm$6.16}}   & {\color[HTML]{000000} 87.95$\pm$6.33}             & \multicolumn{1}{l|}{{\color[HTML]{000000} 79.16$\pm$6.19}} & {\color[HTML]{000000} 87.76$\pm$5.19}             \\ \hline
{\color[HTML]{000000} x-vector}                              & \multicolumn{1}{l|}{{\color[HTML]{000000} {\bf86.43}$\pm$5.35}}   & {\color[HTML]{000000} 91.70$\pm$5.04}             & \multicolumn{1}{l|}{{\color[HTML]{000000} {87.02$\pm$6.98}}} & {\color[HTML]{000000} 91.16$\pm$4.61}             \\ \hline
\end{tabular}
\end{table*}

\begin{figure}
\centering
\begin{tabular}{c}
\large {~~~~~~SVM Classifier} \\
{\includegraphics[width=0.94\textwidth,height=6.5cm]{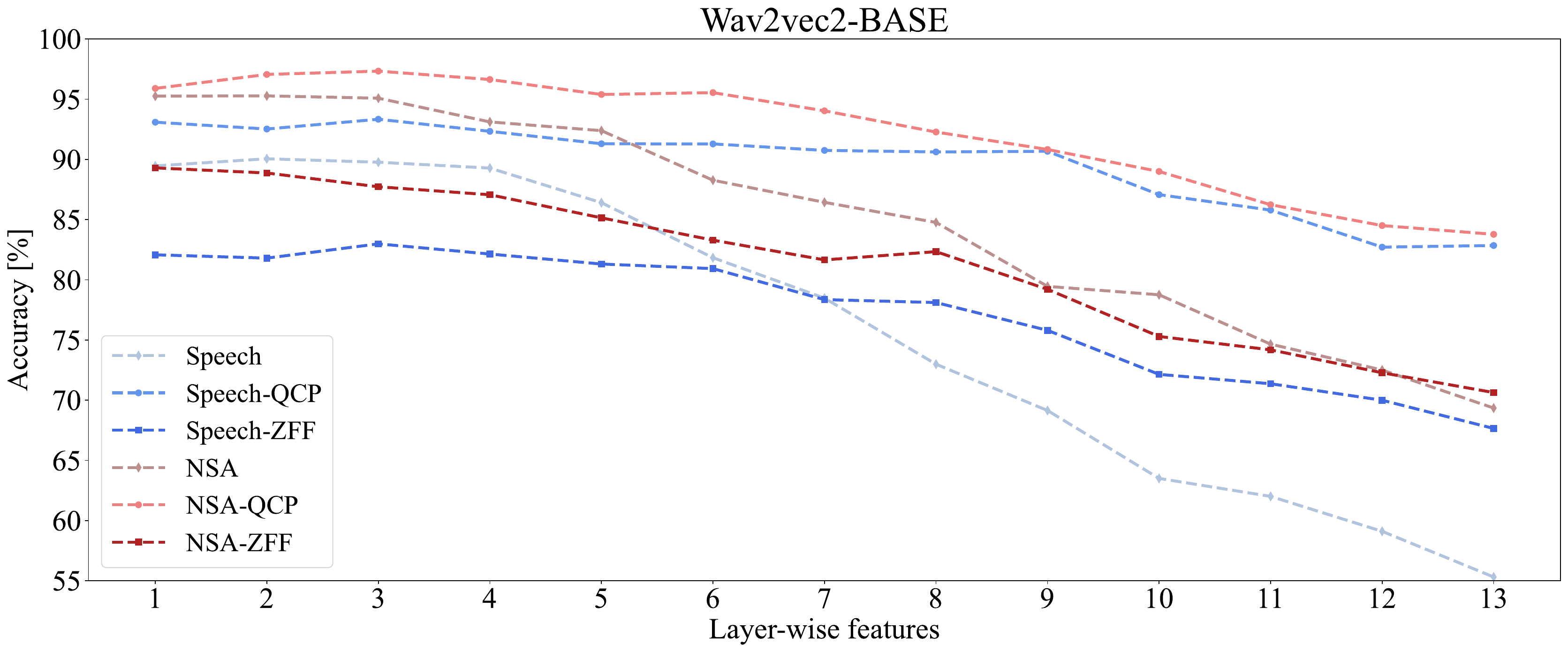}}\\
{\includegraphics[width=0.94\textwidth,height=6.5cm]{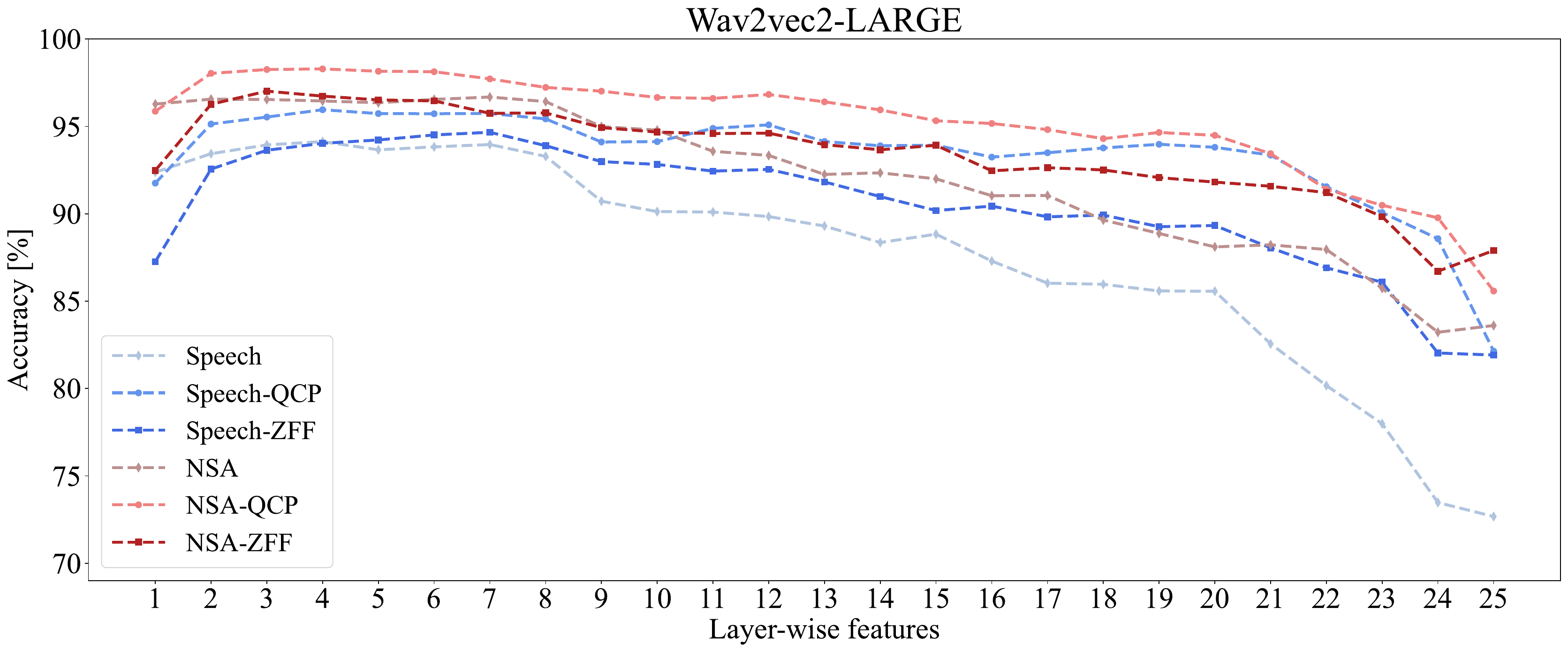}}\\
{\includegraphics[width=0.94\textwidth,height=6.5cm]{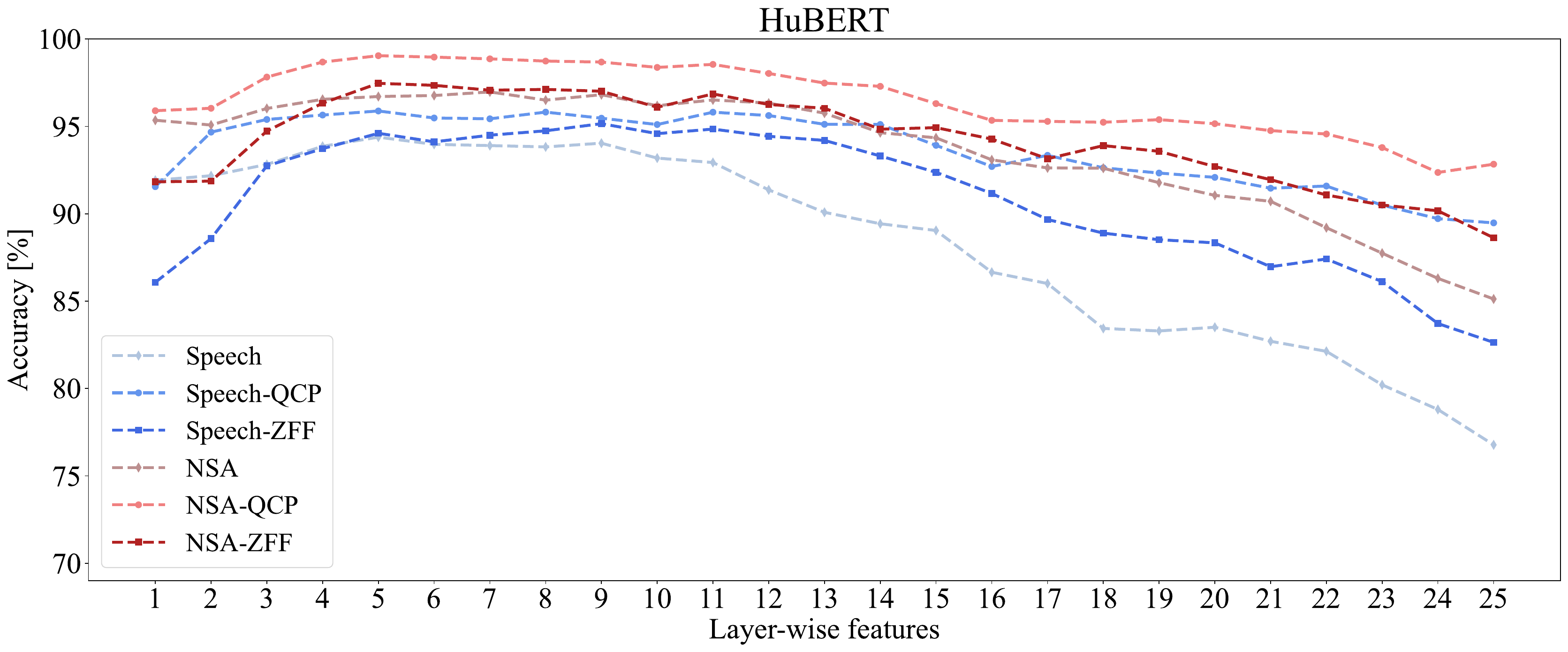}}\\
\end{tabular}
\vspace{-0.5cm}
\caption{{SVM classification accuracy for the features derived from the three pre-trained models: wav2vec2-BASE (top), wav2vec2-LARGE (middle) and HuBERT (bottom). Accuracy is shown for raw signals (speech and NSA) and for glottal source waveforms computed by the QCP and ZFF algorithms from speech (Speech-QCP and Speech-ZFF) and NSA signals (NSA-QCP and NSA-ZFF).}}

\label{fig:pre_trained_layers_svm}
\end{figure}

\begin{figure}
\centering
\begin{tabular}{c}
\large {~~~~~~CNN Classifier} \\
{\includegraphics[width=0.94\textwidth,height=6.5cm]{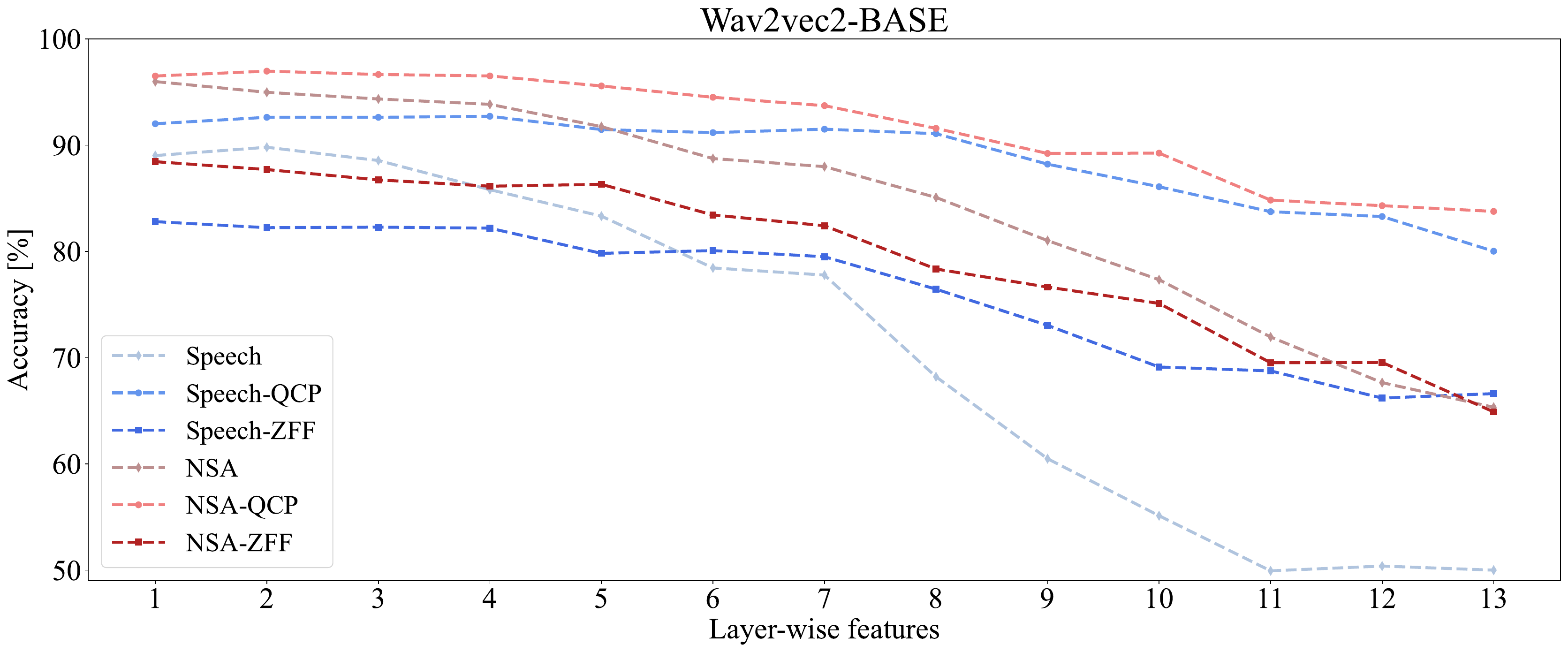}}\\
{\includegraphics[width=0.94\textwidth,height=6.5cm]{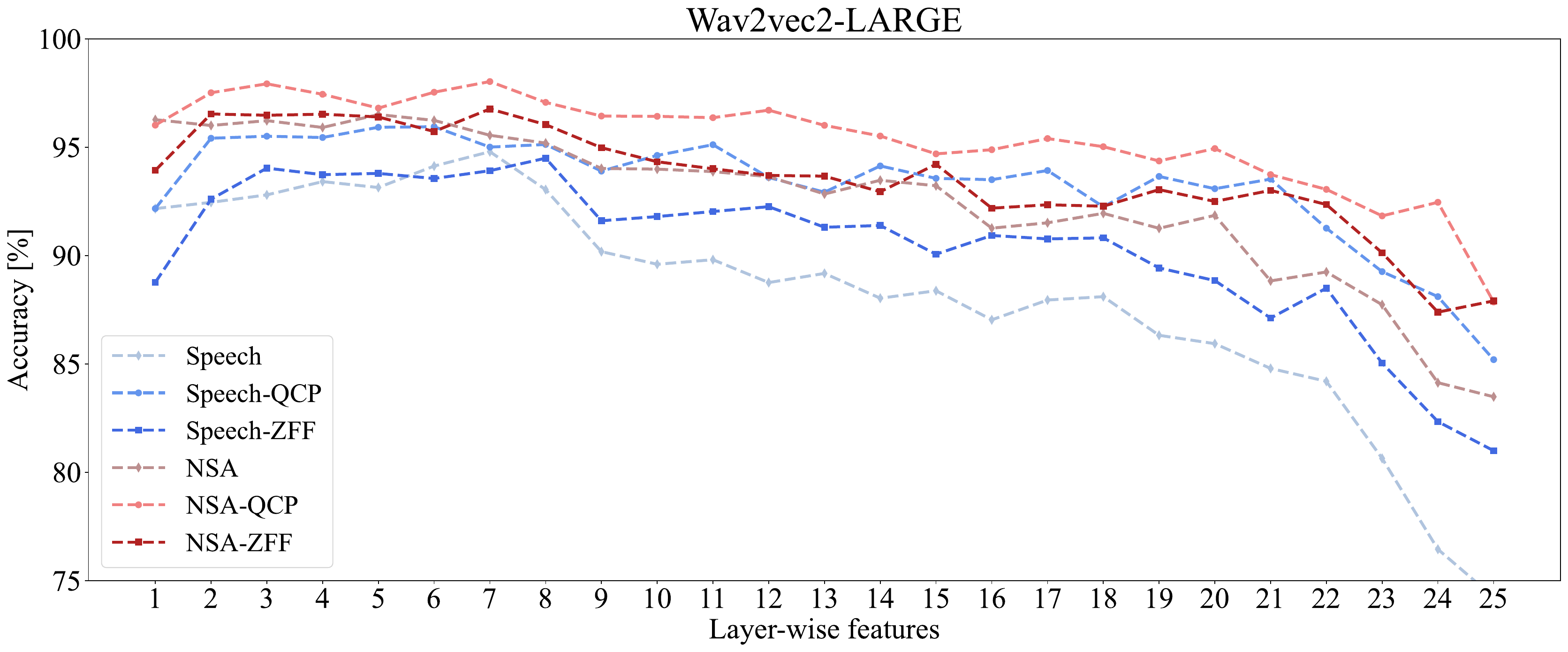}}\\
{\includegraphics[width=0.94\textwidth,height=6.5cm]{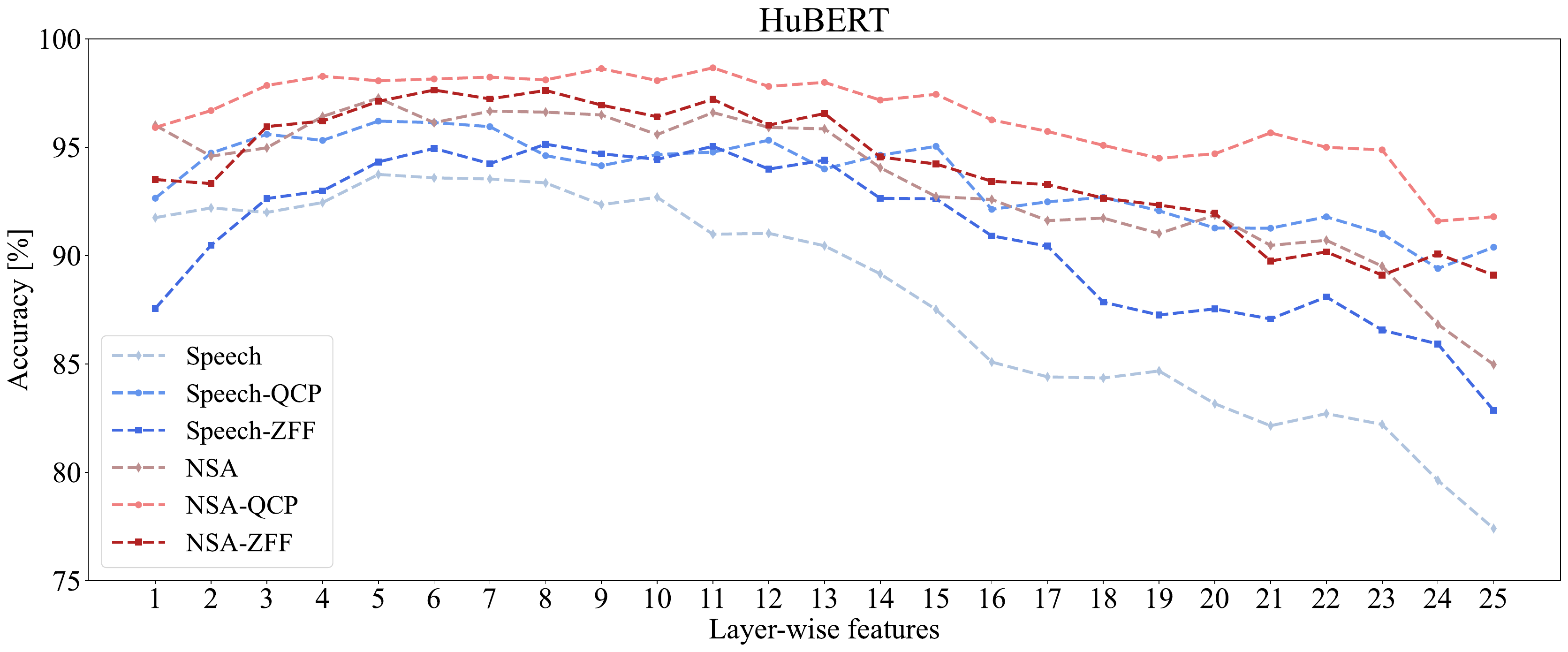}}\\
\end{tabular}
\vspace{-0.5cm}
\caption{{CNN classification accuracy for the features derived from the three pre-trained models: wav2vec2-BASE (top), wav2vec2-LARGE (middle) and HuBERT (bottom). Accuracy is shown for raw signals (speech and NSA) and for glottal source waveforms computed by the QCP and ZFF algorithms from speech (Speech-QCP and Speech-ZFF) and NSA signals (NSA-QCP and NSA-ZFF). 
}
}
\label{fig:pre_trained_layers_cnn}
\end{figure}

\begin{table*}[]
\centering
\vspace{0.1cm}
\caption{{SVM classification accuracy for the conventional features and for the best performing pre-trained model-based features for raw signals (speech and NSA) and their glottal source waveforms (speech-QCP, speech-ZFF, NSA-QCP, and NSA-ZFF).}}
\vspace{0.1cm}

\resizebox{17cm}{3cm}{
\begin{tabular}{|l||l|l|l||l|l|l|}
\hline
\multicolumn{1}{|c||}{{\color[HTML]{000000} Feature}} & \multicolumn{1}{c|}{{\color[HTML]{000000} Speech}} & \multicolumn{1}{c|}{{\color[HTML]{000000} Speech-QCP}} & \multicolumn{1}{c||}{{\color[HTML]{000000} Speech-ZFF}} & \multicolumn{1}{c|}{{\color[HTML]{000000} NSA}} & \multicolumn{1}{c|}{{\color[HTML]{000000} NSA-QCP}} & \multicolumn{1}{c|}{{\color[HTML]{000000} NSA-ZFF}} \\ \hline \hline

\multicolumn{7}{|c|}{\bf Conventional features}\\ \hline \hline

{\color[HTML]{000000} Spectrogram}                & {\color[HTML]{000000} 81.72$\pm$6.09}                & {\color[HTML]{000000} 76.86$\pm$8.85}                    & {\color[HTML]{000000} 69.96$\pm$8.40}                    & {\color[HTML]{000000} 91.38$\pm$4.74}             & {\color[HTML]{000000} 89.70$\pm$5.37}                 & {\color[HTML]{000000} 80.18$\pm$7.41}                 \\ \hline
{\color[HTML]{000000} Mel-spectrogram}                & {\color[HTML]{000000} 83.99$\pm$6.73}                & {\color[HTML]{000000} 72.46$\pm$8.96}                    & {\color[HTML]{000000} 74.33$\pm$6.69}                    & {\color[HTML]{000000} 92.84$\pm$4.99}             & {\color[HTML]{000000} 87.25$\pm$6.26}                 & {\color[HTML]{000000} 81.89$\pm$8.10}                 \\ \hline
{\color[HTML]{000000} MFCC       }                & {\color[HTML]{000000} 79.78$\pm$6.79}                & {\color[HTML]{000000} 80.42$\pm$6.67}                    & {\color[HTML]{000000} 79.45$\pm$6.88}                    & {\color[HTML]{000000} 89.01$\pm$5.09}             & {\color[HTML]{000000} 82.04$\pm$7.92}                 & {\color[HTML]{000000} 86.55$\pm$6.21}                 \\ \hline
{\color[HTML]{000000} i-vector       }                & {\color[HTML]{000000} 81.05$\pm$6.16}                & {\color[HTML]{000000} 86.90$\pm$5.93}                    & {\color[HTML]{000000} 88.80$\pm$5.74}                    & {\color[HTML]{000000} 87.95$\pm$6.33}             & {\color[HTML]{000000} 93.72$\pm$4.48}                 & {\color[HTML]{000000} 94.24$\pm$4.55}                 \\ \hline
{\color[HTML]{000000} x-vector       }                & {\color[HTML]{000000} 86.43$\pm$5.35}                & {\color[HTML]{000000} 89.60$\pm$6.50}                    & {\color[HTML]{000000} {\bf91.17}$\pm$5.48}                    & {\color[HTML]{000000} 91.70$\pm$5.04}             & {\color[HTML]{000000} {\bf95.90}$\pm$2.86}                 & {\color[HTML]{000000} 94.13$\pm$4.16}                 \\ \hline
\multicolumn{7}{|c|}{\bf Features based on pre-trained models}\\ \hline \hline

{\color[HTML]{000000} wav2vec2-BASE}                  & {\color[HTML]{000000} 90.05$\pm$6.11}                & {\color[HTML]{000000} 93.33$\pm$4.14}                    & {\color[HTML]{000000} 82.98$\pm$6.09}                    & {\color[HTML]{000000} 95.27$\pm$3.41}             & {\color[HTML]{000000} 97.33$\pm$2.19}                 & {\color[HTML]{000000} 89.29$\pm$5.17}                 \\ \hline
{\color[HTML]{000000} wav2vec2-LARGE}                 & {\color[HTML]{000000} 94.13$\pm$4.23}                & {\color[HTML]{000000} {\bf95.95}$\pm$3.29}                    & {\color[HTML]{000000} 94.66$\pm$3.66}                    & {\color[HTML]{000000} 96.68$\pm$3.58}             & {\color[HTML]{000000} 98.29$\pm$2.17}                 & {\color[HTML]{000000} 97.02$\pm$3.03}                 \\ \hline
{\color[HTML]{000000} HuBERT}                         & {\color[HTML]{000000} 94.38$\pm$3.71}                & {\color[HTML]{000000} 95.88$\pm$4.11}                    & {\color[HTML]{000000} 95.15$\pm$4.02}                    & {\color[HTML]{000000} 96.97$\pm$3.0}              & {\color[HTML]{000000} {\bf99.04}$\pm$2.08}                 & {\color[HTML]{000000} 97.46$\pm$2.59}                 \\ \hline
\end{tabular}}
\label{tab:pre_trained_layers_svm1}
\end{table*}


\begin{table*}
\centering
\vspace{0.1cm}
\caption{{CNN classification accuracy for the conventional features and for the best performing pre-trained model-based features for raw signals (speech and NSA) and their glottal source waveforms (speech-QCP, speech-ZFF, NSA-QCP, and NSA-ZFF).}}
\vspace{0.1cm}

\resizebox{17cm}{3cm}{
\begin{tabular}{|l||l|l|l||l|l|l|}
\hline
\multicolumn{1}{|c||}{{\color[HTML]{000000} Feature}} & \multicolumn{1}{c|}{{\color[HTML]{000000} Speech}} & \multicolumn{1}{c|}{{\color[HTML]{000000} Speech-QCP}} & \multicolumn{1}{c||}{{\color[HTML]{000000} Speech-ZFF}} & \multicolumn{1}{c|}{{\color[HTML]{000000} NSA}} & \multicolumn{1}{c|}{{\color[HTML]{000000} NSA-QCP}} & \multicolumn{1}{c|}{{\color[HTML]{000000} NSA-ZFF}} \\ \hline \hline

\multicolumn{7}{|c|}{\bf Conventional features}\\ \hline \hline

{\color[HTML]{000000} Spectrogram}                & {\color[HTML]{000000} 86.18$\pm$6.58}                & {\color[HTML]{000000} 80.26$\pm$8.23}                    & {\color[HTML]{000000} 81.67$\pm$7.33}                    & {\color[HTML]{000000} 92.19$\pm$5.86}             & {\color[HTML]{000000} {\bf95.25}$\pm$4.13}                 & {\color[HTML]{000000} 87.67$\pm$5.52}                 \\ \hline
{\color[HTML]{000000} Mel-spectrogram}                & {\color[HTML]{000000} 87.74$\pm$5.65}                & {\color[HTML]{000000} 79.07$\pm$7.34}                    & {\color[HTML]{000000} 78.39$\pm$7.51}                    & {\color[HTML]{000000} 92.76$\pm$4.84}             & {\color[HTML]{000000} 95.02$\pm$4.35}                 & {\color[HTML]{000000} 86.81$\pm$5.92}                 \\ \hline
{\color[HTML]{000000} MFCC       }                & {\color[HTML]{000000} 77.93$\pm$6.41}                & {\color[HTML]{000000} 77.96$\pm$6.92}                    & {\color[HTML]{000000} 77.59$\pm$6.71}                    & {\color[HTML]{000000} 87.36$\pm$4.65}             & {\color[HTML]{000000} 83.31$\pm$9.02}                 & {\color[HTML]{000000} 86.78$\pm$6.09}                 \\ \hline
{\color[HTML]{000000} i-vector       }                & {\color[HTML]{000000} 79.16$\pm$6.19}                & {\color[HTML]{000000} 85.59$\pm$5.60}                    & {\color[HTML]{000000} 89.79$\pm$5.08}                    & {\color[HTML]{000000} 87.76$\pm$5.19}             & {\color[HTML]{000000} 93.15$\pm$4.13}                 & {\color[HTML]{000000} 94.30$\pm$3.96}                 \\ \hline
{\color[HTML]{000000} x-vector       }                & {\color[HTML]{000000} 87.02$\pm$6.98}                & {\color[HTML]{000000} 87.71$\pm$6.34}                    & {\color[HTML]{000000} {\bf90.54}$\pm$6.12}                    & {\color[HTML]{000000} 91.16$\pm$4.61}             & {\color[HTML]{000000} {\bf95.08}$\pm$3.91}                 & {\color[HTML]{000000} 93.43$\pm$3.81}                 \\ \hline
\multicolumn{7}{|c|}{\bf Features based on pre-trained models}\\ \hline \hline

{\color[HTML]{000000} wav2vec2-BASE}                  & {\color[HTML]{000000} 89.81$\pm$6.39}                & {\color[HTML]{000000} 92.73$\pm$5.11}                    & {\color[HTML]{000000} 82.81$\pm$6.98}                    & {\color[HTML]{000000} 95.99$\pm$4.74}             & {\color[HTML]{000000} 96.97$\pm$3.18}                 & {\color[HTML]{000000} 88.45$\pm$6.3}                  \\ \hline
{\color[HTML]{000000} wav2vec2-LARGE}                 & {\color[HTML]{000000} 94.8$\pm$3.36}                 & {\color[HTML]{000000} 95.95$\pm$3.33}                    & {\color[HTML]{000000} 94.5$\pm$4.69}                     & {\color[HTML]{000000} 96.51$\pm$3.26}             & {\color[HTML]{000000} 98.04$\pm$2.45}                 & {\color[HTML]{000000} 96.77$\pm$2.98}                 \\ \hline
{\color[HTML]{000000} HuBERT}                         & {\color[HTML]{000000} 93.75$\pm$4.42}                & {\color[HTML]{000000} {\bf96.21}$\pm$3.23}                    & {\color[HTML]{000000} 95.15$\pm$4.29}                    & {\color[HTML]{000000} 97.27$\pm$2.48}             & {\color[HTML]{000000} {\bf98.67}$\pm$2.33}                 & {\color[HTML]{000000} 97.64$\pm$2.7}                  \\ \hline
\end{tabular}}
\label{tab:pre_trained_layers_cnn}
\end{table*}

Table~\ref{Tab:Result_nsa_speech_conventional} shows the results for the five baseline features (in terms of the mean and standard deviation of accuracy) for the speech and NSA signals using the SVM and CNN classifiers. From the table, it can be clearly seen that the classification accuracy is higher for the NSA signal compared to the speech signal for all five features (spectrogram, mel-spectrogram, MFCC, i-vector, and x-vector) and for both classifiers. This implies that, in comparison to the acoustic speech signal, the NSA signals contains a larger amount of information about voice quality due to its better correspondence with vocal fold vibration patterns which occur when the speaker changes voice quality. 
{Among the conventional features, the mel-spectrogram performed better than the other four features (except for the x-vector for the speech signal using the SVM classifier). 
For a given feature and input modality combination, there were in general small differences in accuracy between the two classifiers (except that CNN was better than SVM in using the spectrogram or mel-spectrogram feature for the speech signal).}

{Figure~\ref{fig:pre_trained_layers_svm} shows the results (in terms of mean accuracy) for the features derived from the three pre-trained models (wav2vec2-BASE, wav2vec2-LARGE and HuBERT) for raw signals (speech and NSA) and for glottal source waveforms using the SVM classifier. Similarly, Fig.~\ref{fig:pre_trained_layers_cnn} shows the results for the CNN classifier. From these figures, it can be clearly seen that the features derived from all the pre-trained models (especially for the initial layers) for raw signals (speech and NSA) and glottal source waveforms performed better than the conventional features (see Tables~\ref{Tab:Result_nsa_speech_conventional}-~\ref{tab:pre_trained_layers_cnn}). In all the pre-trained models, it can be clearly seen that the features derived from the initial layers showed better performance compared to those derived from the later layers. This suggests that the initial layers of the pre-trained models have learned generic features that can be effectively used for the classification of voice qualities.
It can also be observed that classification accuracy is higher for the NSA signal and their glottal source waveforms (NSA-QCP and NSA-ZFF) compared to the speech and their glottal source waveforms (Speech-QCP and Speech-ZFF). This observation indicates that NSA contains better information about voice quality due to its capability of capturing vocal fold vibration patterns.}

Results for the conventional reference features and for the best performing features from the pre-trained models are given for raw signals (speech and NSA) and glottal source waveforms (speech-QCP, speech-ZFF, NSA-QCP, and NSA-ZFF) { in Tables~\ref{tab:pre_trained_layers_svm1} and~\ref{tab:pre_trained_layers_cnn}} for the SVM and CNN classifier, respectively. From the tables, it can be clearly seen that the features based on the pre-trained models show better performance compared to the conventional features for both raw signals and their glottal source waveforms. Between raw signals and glottal source waveforms, glottal source waveforms show better performance in many cases. Between the two glottal source waveform estimation methods, the QCP method shows better performance compared to the ZFF method for the majority of the cases. Among the conventional features, the x-vector feactures performed better than the other features in most cases. Among the three pre-trained models, it can be seen that the HuBERT-based features gave better performance compared to the wav2vec2-BASE and wav2vec2-LARGE-based features. 

\begin{figure*}[ht]
\centering
\includegraphics[width=\columnwidth,height=8cm,trim={0.9cm 0cm 0.7cm 0cm},clip]{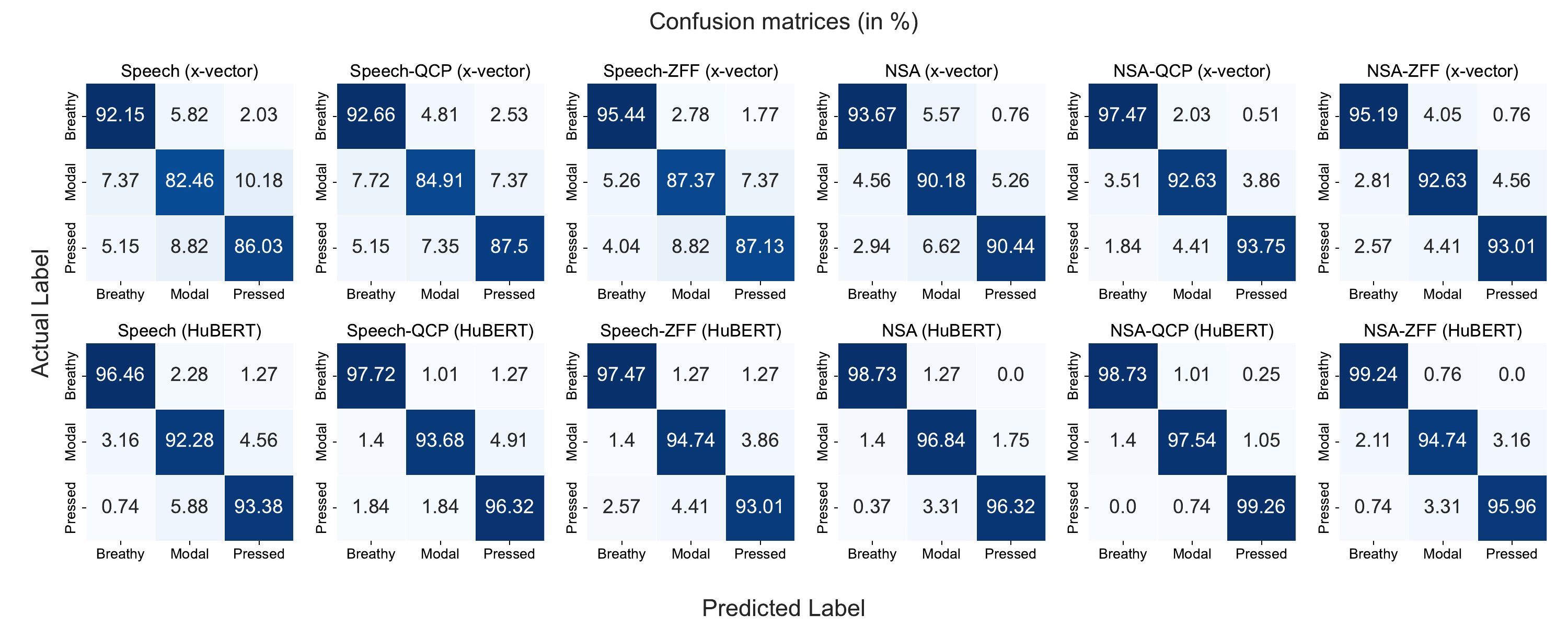}
\vspace{-0.4cm}
\caption{\label{fig:confusion_matrix} {Confusion matrices  for the x-vector (top row) and for the HuBERT features (bottom row) for raw signals (speech and NSA) and their glottal source waveforms using the CNN classifier.}}
\end{figure*}

\begin{figure*}
\centering
\begin{tabular}{ccc}
{\bf{~~~~~~~~~~Speech-QCP(x-vector)}} & {\bf{~~~~~~~~~~NSA-QCP(x-vector)}}\\
\includegraphics[width=8.1cm,trim={0cm 0cm 0cm 1.1cm},clip]{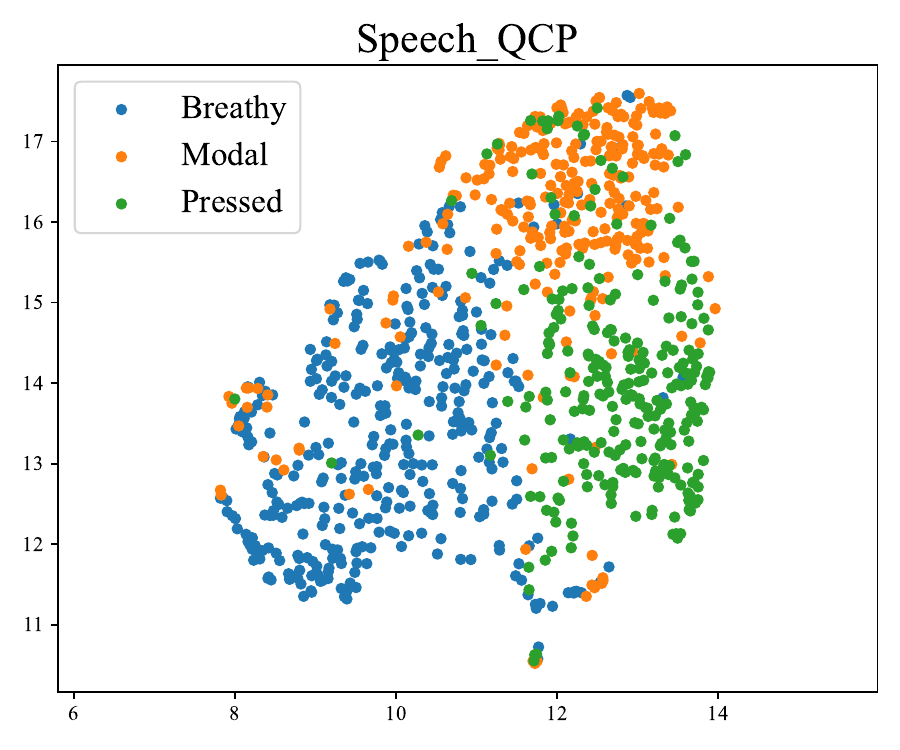}&
\includegraphics[width=8.1cm,trim={0cm 0cm 0cm 1cm},clip]{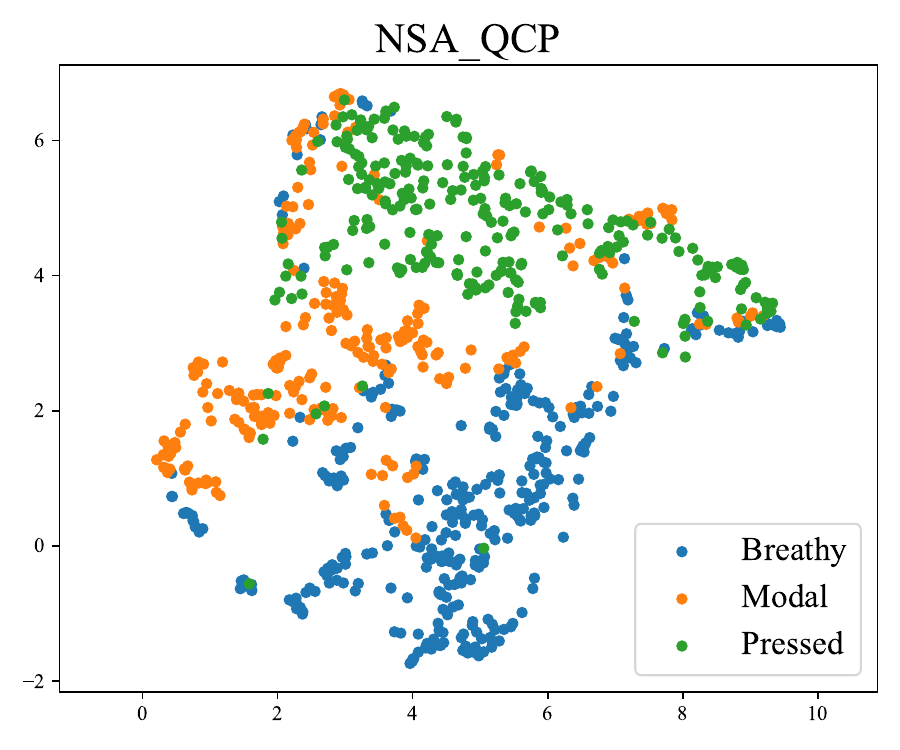}\\
{\bf{~~~~~~~~~~Speech-QCP(HuBERT)}} & {\bf{~~~~~~~~~~NSA-QCP(HuBERT)}}\\

\includegraphics[width=8.1cm,trim={0cm 0cm 0cm 1.1cm},clip]{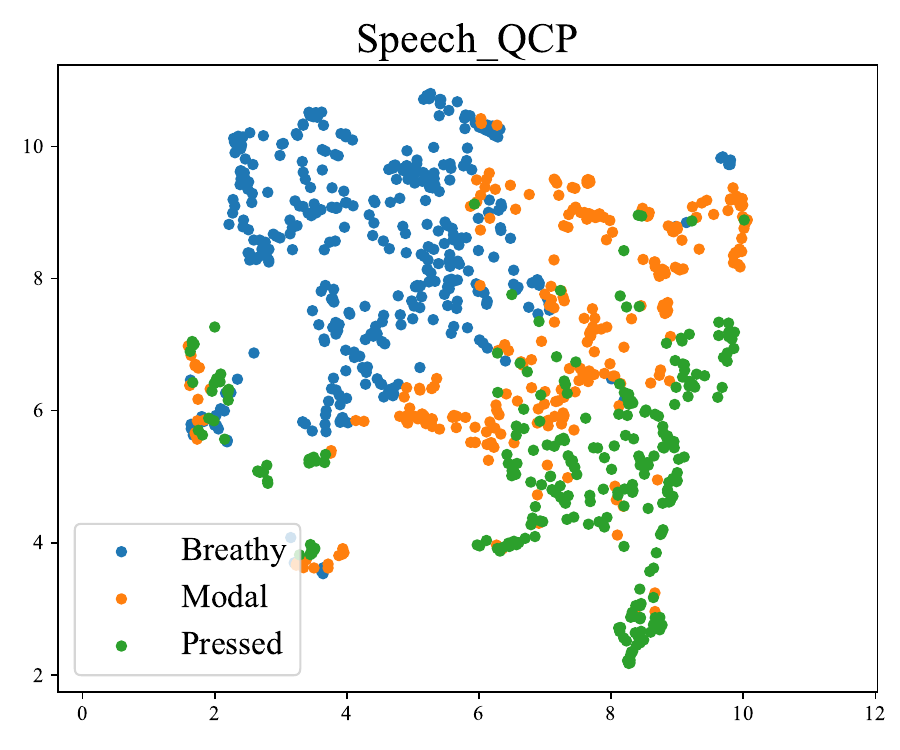}&

\includegraphics[width=8.1cm,trim={0cm 0cm 0cm 1cm},clip]{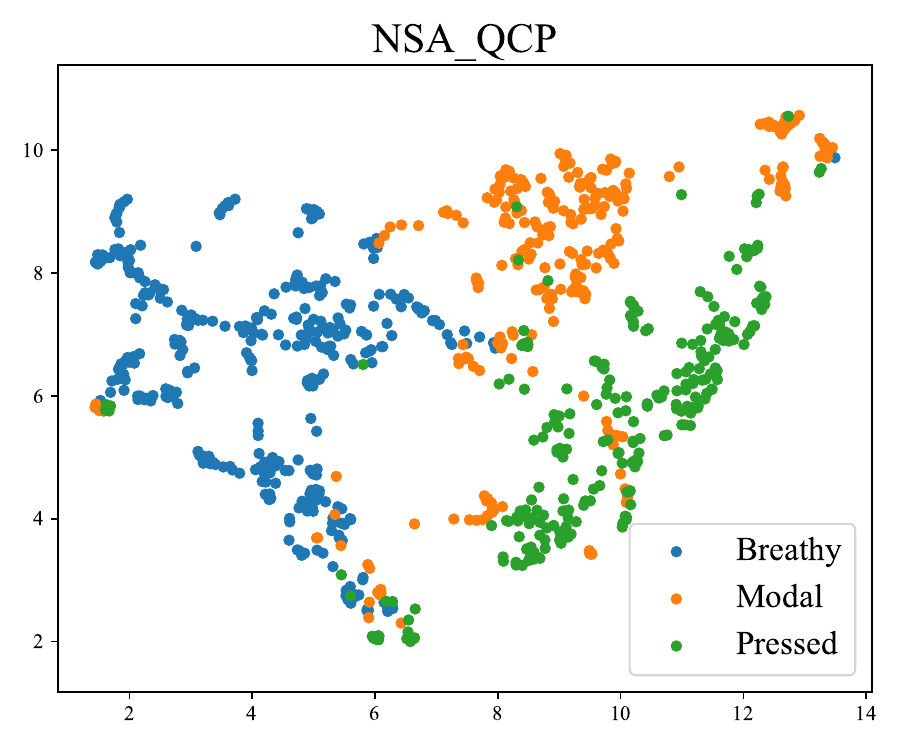}\\
\end{tabular}
\caption{ Visualizations using the UMAP algorithm for the x-vector (top row) and the HuBERT feature representations (bottom row) of the speech-QCP and NSA-QCP signals.}
\label{fig:umaps}
\end{figure*}

{In the case of SVM, the x-vector features were best among the conventional features for speech (speech-ZFF: 91.17\%) and NSA (NSA-QCP: 95.90\%) signals. With the same classifier, however, the wav2vec2-LARGE and HuBERT-based features gave better performance for the glottal source waveforms derived using the QCP method for both speech and NSA signals (i.e., speech-QCP: 95.95\% and NSA-QCP: 99.04\%). Similarly, in the case of CNN, the x-vector was the best conventional feature for speech signals (speech-ZFF: 90.54\%), and the spectrogram feature was the best conventional feature for NSA signals (NSA-QCP: 95.25\%). For the same classifier utilizing the pre-trained model-based features, the HuBERT-based features gave better performance for the glottal source waveforms derived using the QCP method for both speech and NSA signals (i.e., speech-QCP: 96.21\% and NSA-QCP: 98.67\%). Overall, the HuBERT-based features showed an absolute accuracy improvement of 3\%--6\% for speech and NSA signals compared to the conventional features.

Confusion matrices are displayed in Fig.~\ref{fig:confusion_matrix} for the x-vector (top row) and for the HuBERT feature (bottom row) for raw signals (speech and NSA) and for glottal source waveforms based on the CNN classifier. It can be clearly seen that there are fewer confusions between the three voice qualities for the HuBERT features compared to the x-vector in both raw signals and glottal source waveforms.

Finally, we employed the UMAP (Uniform Manifold Approximation and Projection) algorithm \cite{mcinnes2018umap} to visually examine the capability of the latent representations (features) to discriminate the three voice qualities. The UMAP algorithm reduces the dimensionality of the latent features into a lower-dimensional space. Figure~\ref{fig:umaps} displays the UMAP visualizations for the x-vector (top row) and the HuBERT feature representations (bottom row) of the speech-QCP and NSA-QCP signals. It can be observed that all the three voice qualities (breathy, modal, and pressed) are well clustered in both of the feature representations 
and for both the speech-QCP and NSA-QCP signals. The clusters in the NSA-QCP signal appear to be more distinct from each other compared to the clusters in the speech-QCP signal. Moreover, the clusters in the HuBERT feature representations are better separated compared to those in the x-vector. These observations are in line with  the trends observed in the accuracy results and confusion matrices reported in Tables~\ref{tab:pre_trained_layers_svm1} and~\ref{tab:pre_trained_layers_cnn}, and Fig.~\ref{fig:confusion_matrix}.} 

\section{Summary and Conclusions}
In this article, the automatic classification of voice quality was studied between three classes (breathy, modal, pressed). The classification task was studied using the simultaneously-recorded raw signals (speech and NSA) and their glottal source waveforms. From these input signals, features were derived using three self-supervised pre-trained models (wav2vec2-BASE, wav2vec2-LARGE, and HuBERT) and using SVM and CNN as classifiers. Our results indicated that the NSA signal shows a better capability in the classification task compared to the acoustical speech signal. In addition, the experiments showed that the features derived from the self-supervised pre-trained models were clearly more effective in the voice quality classification compared to five conventional features (spectrogram, mel-spectrogram, MFCCs, i-vector, and x-vector). Furthermore, the features derived from the pre-trained models using the glottal source waveform as input were more effective in comparison to the same features extracted from raw signals (in both speech and NSA). The two classifiers performed equally well for all the pre-trained model-based features for both speech and NSA signals, and their glottal source waveforms. In all the three pre-trained models, the features derived from the initial layers performed better than the features from the last layers. This is because the initial layers contain information about phones, which is useful in voice quality classification. On the other hand, the layers that are located closer to the final layer carry information about the phoneme identity (because the models have been fine-tuned for ASR), and hence they may not be effective in the voice quality classification task. Among the pre-trained models, it was found that the HuBERT-based features performed better than the wav2vec2-BASE and wav2vec2-LARGE features, for both speech and NSA signals, and their glottal source waveforms. { In particular, the HuBERT-based features showed an absolute accuracy improvement of 3\%--6\% for speech and NSA signals in the classification of voice qualities in comparison to the conventional features.} 

In conclusion, the current study shows encouraging evidence according to which the recently developed self-supervised pre-trained models (wav2vec2 and HuBERT) provide highly effective tools to extract features for the automatic classification of voice qualities. The performance of the classification is further improved by utilizing the NSA signal as input instead of the acoustical speech signal, and by removing the effect of vocal tract resonances using inverse filtering. 

\section*{Acknowledgements}
This study was supported by the Academy of Finland (project no. 330139) and Aalto University (the Ministry of Education and Culture program for India). Aalto Science IT provided the computational resources.


\end{document}